\documentclass{article}
\usepackage{spconf,amsmath,graphicx,wrapfig,bm,subcaption, multirow}
\usepackage{amsfonts,amssymb}
\usepackage{fancyhdr}

\title{TF-SepNet: An Efficient 1D Kernel Design in CNNs for Low-Complexity Acoustic Scene Classification}
\name{Yiqiang Cai, Peihong Zhang, Shengchen Li}
\address{School of Advanced Technology\\
Xi'an Jiaotong-Liverpool University\\
111 Ren'ai Road, Suzhou, China }
\begin{document}
\maketitle
\thispagestyle{fancy}
\fancyhead{}
\lhead{}
\lfoot{\fbox{\parbox[t]{\textwidth}{\copyright2023 IEEE. Personal use of this material is permitted. Permission from IEEE must be obtained for all other uses, in any current or future media, including reprinting/republishing this material for advertising or promotional purposes, creating new collective works, for resale or redistribution to servers or lists, or reuse of any copyrighted component of this work in other works.}}}
\begin{abstract}
Recent studies focus on developing efficient systems for acoustic scene classification (ASC) using convolutional neural networks (CNNs), which typically consist of consecutive kernels. This paper highlights the benefits of using separate kernels as a more powerful and efficient design approach in ASC tasks. Inspired by the time-frequency nature of audio signals, we propose TF-SepNet, a CNN architecture that separates the feature processing along the time and frequency dimensions. Features resulted from the separate paths are then merged by channels and directly forwarded to the classifier. Instead of the conventional two dimensional (2D) kernel, TF-SepNet incorporates one dimensional (1D) kernels to reduce the computational costs. Experiments have been conducted using the TAU Urban Acoustic Scene 2022 Mobile development dataset. The results show that TF-SepNet outperforms similar state-of-the-arts that use consecutive kernels. A further investigation reveals that the separate kernels lead to a larger effective receptive field (ERF), which enables TF-SepNet to capture more time-frequency features.
\end{abstract}
\begin{keywords}
Acoustic scene classification, efficient neural networks, separated kernels, effective receptive field
\end{keywords}

% \blfootnote{\copyright2023 IEEE. Personal use of this material is permitted. Permission from IEEE must be obtained for all other uses, in any current or future media, including reprinting/republishing this material for advertising or promotional purposes, creating new collective works, for resale or redistribution to servers or lists, or reuse of any copyrighted component of this work in other works.}
\section{Introduction}
\label{sec:intro}
Acoustic scene classification (ASC) \cite{barchiesi2015acoustic} is a basic audio processing task that identifies and classifies audio signals into predefined environmental sound scenes such as airports, parks and urban streets. ASC systems usually require a delicate balance between accuracy and computational efficiency, especially when aiming for real-time processing and deployment on resource-constrained devices \cite{MartinMorato2022}.

\begin{figure}
\centering
\includegraphics[width=0.7\linewidth]{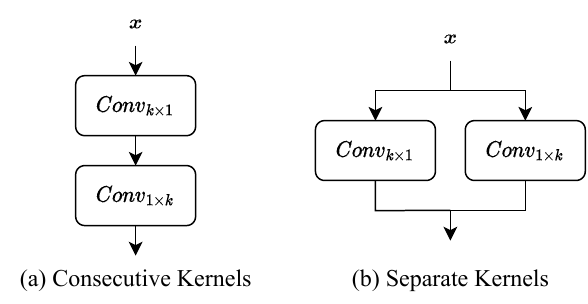} 
\caption{\textbf{Simplified diagrams} of two 1D-kernel-based design approaches in CNNs. $x$ denotes the input features.}
\label{fig:kernels}
\end{figure}

In response to the demand for efficient ASC systems, researchers predominantly focus on harnessing the power of convolutional neural networks (CNNs) \cite{abesser2020review}. Traditional approaches develop CNN-based ASC systems by stacking multiple two-dimensional (2D) kernels \cite{MartinMorato2022,Koutini2019,Hu2021}, while recent studies \cite{Cho2019,kim21l_interspeech,Lee2022,Phan2022} have explored the use of one-dimensional (1D) kernel as a potential alternative. The 1D kernel is initially introduced to address the overfitting problem in ASC tasks \cite{Cho2019}, which uses two consecutive 1D kernels with size of $k \times 1$ and $1 \times k$ as depicted in (a) of Fig. \ref{fig:kernels}. It is worth noting that the number of parameters of two 1D ($k \times 1$ and $1 \times k$) kernels is less than that of a 2D $k \times k$ kernel. Therefore, the consecutive 1D kernels are then employed by subsequent studies \cite{kim21l_interspeech,Lee2022,Phan2022} for the design of efficient ASC models.

\begin{figure*}[t]
\centering
\includegraphics[width=\linewidth]{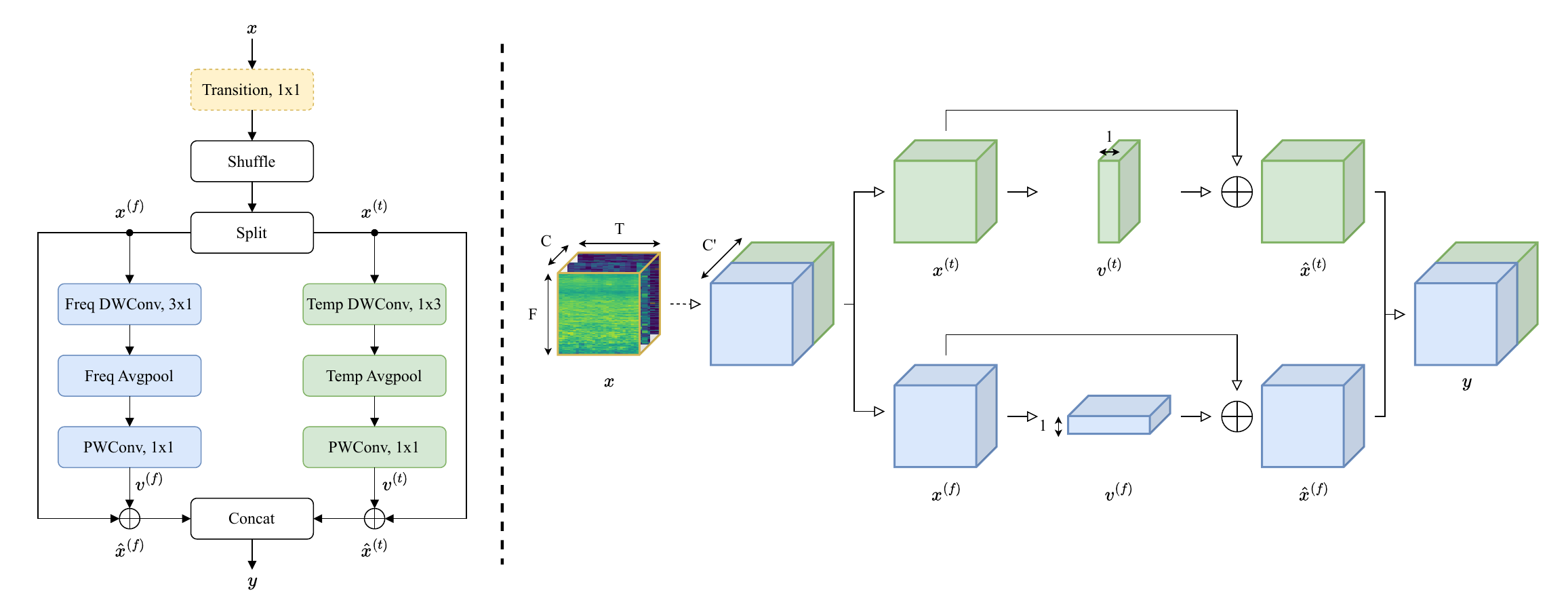} 
\caption{\textbf{Left:} Visualization of the Time-Frequency Separate Convolutions (TF-SepConvs) module. \textbf{Right:} Transformation of features maps. \textbf{DWConv} represents the depthwise convolution and \textbf{PWConv} denotes the pointwise convolution. \textbf{Freq/Temp} prefix denotes the 1D operation on the frequential or temporal axis. \textbf{Shuffle} indicates the channel shuffle unit \cite{Zhang2018}. The dashed line or box means corresponding operation only exists when the channel number changes. The input feature $x$ is in $\mathbb{R}^{C \times F \times T}$, where $C, F, T$ respectively denotes channel, frequency and time dimensions.}
\label{fig:tfsp}
\end{figure*}

The inspiration behind this paper stems from the time-frequency nature of audio signals. A recent work \cite{Mu2021} introduces Harmonic-Percussive Source Separation (HPSS) to disentangle the time and frequency information from the audio spectrograms, which effectively helps CNNs to learn crucial time–frequency features. However, the additional HPSS preprocessing requires considerable computational cost. Instead of appending a preprocessing component, we apply the principle of disentanglement to design a novel CNN architecture, Time-Frequency Separate Network (TF-SepNet). Specifically, our proposed network incorporates two independent paths for extracting features along the time and frequency dimensions. As illustrated in (b) of Fig. \ref{fig:kernels}, the $k \times 1$ kernel is responsible for extracting frequency features, while the $1 \times k$ kernel focuses on temporal features. The resulted features are then concatenated in the channel dimension and directly forwarded to the classifier. Our network generally adheres to the macro architecture of BC-ResNet \cite{kim21l_interspeech}, with the integration of depthwise separable convolution \cite{Phan2022} and broadcasting operation \cite{kim21l_interspeech} to further reduce the model complexity. 

To evaluate the effectiveness of TF-SepNet, experiments were conducted using the TAU Urban Acoustic Scene 2022 Mobile development dataset \cite{Mesaros2018}, a benchmark widely used in the ASC research community. The results reveal that TF-SepNet gets higher accuracy than similar approaches of consecutive kernels \cite{kim21l_interspeech,Lee2022} while exhibiting lower computational complexity. To examine the rationale behind the improvements observed in TF-SepNet, we conduct an in-depth analysis of Effective Receptive Fields (ERF) \cite{luo2016understanding} for the models being compared. The ERF refers to the region within input features that a particular neuron is sensitive to. Achieving a larger receptive field typically involves the adjustment of hyperparameters in CNNs such as kernel size and down-sampling layers \cite{Koutini2019}. Our investigation reveals that the design of separate kernels in TF-SepNet results in a notably larger ERF compared to the approaches that use consecutive kernels. We believe the larger ERF of TF-SepNet substantially improves the model's capacity to capture vital time-frequency features within acoustic scene sounds.

\section{Proposed Method}
\label{sec:method}

\subsection{Time-Frequency Separate Convolutions}
\label{ssec:tf-sep}
Inspired by the time-frequency attention mechanism \cite{Mu2021}, our study introduces a novel 1D-kernel-based module, Time-Frequency Separate Convolutions (TF-SepConvs), for separating the processing of feature maps in CNNs. Different from the approaches presented in \cite{kim21l_interspeech,Lee2022,Phan2022}, the 1D kernels in TF-SepConvs independently deal with two halves of feature maps so as to capture distinct information in the time and frequency dimensions. The consistency of channel numbers is ensured by the division and subsequent concatenation of feature maps in the channel dimension. In addition, depthwise separable convolution \cite{Phan2022} and broadcasting operation \cite{kim21l_interspeech} are integrated into TF-SepConvs for further reducing the parameters and computational overheads. To offer a comprehensive understanding of the structure of TF-SepConvs, we provide a detailed explanation below.

As illustrated in Fig. \ref{fig:tfsp}, TF-SepConvs begin with a transition layer, consisting of a $1 \times 1$ convolution. The transition layer serves the purpose of expanding or shrinking the number of channels from $C$ to $C'$, $x\in\mathbb{R}^{C \times F \times T}\rightarrow\mathbb{R}^{C '\times F \times T}$. After the transition layer, a shuffle unit \cite{Zhang2018} is introduced to establish connections of the feature maps between channels. Following the shuffle unit, the feature maps are evenly divided into two halves by channels: $x^{(f)},x^{(t)}\in\mathbb{R}^{C'/2 \times F \times T}$. $x^{(f)}$ and $x^{(t)}$ are then separately processed by operations in the frequential path and temporal path. As shown in equation (\ref{eq:avgf}) (\ref{eq:avgt}), these two paths consist of a ($3 \times 1$ or $1 \times 3$) depthwise convolution denoted as $d_{3\times 1}$ or $d_{1\times 3}$, an (frequency or time) average pool and a $1 \times 1$ pointwise convolution denoted as $p_{1\times 1}$, where $v^{(f)}\in\mathbb{R}^{C'/2 \times 1 \times T}$ and $v^{(t)}\in\mathbb{R}^{C'/2 \times F \times 1}$. All convolutions mentioned above are followed by batch normalization (BN) and relu activation (ReLu).

\begin{equation}
\label{eq:avgf}
    v^{(f)} = p_{1\times1}(\frac{1}{F}\sum\nolimits_{i=1}^{F}d_{3\times1}(x_{ij}^{(f)}))
\end{equation}

\begin{equation}
\label{eq:avgt}
    v^{(t)} = p_{1\times1}(\frac{1}{T}\sum\nolimits_{j=1}^{T}d_{1\times3}(x_{ij}^{(t)}))
\end{equation}

The 1D features $v^{(f)}$ and $v^{(t)}$ are then respectively expanded to 2D shape as shown in equation (\ref{eq:bcf}) (\ref{eq:bct}), where $\hat{x}^{(f)},\hat{x}^{(t)}\in\mathbb{R}^{C'/2 \times F \times T}$. The process of averaging and expanding is known as the broadcasting operation \cite{kim21l_interspeech}. The broadcasting operation effectively  decreases the size of feature maps for the pointwise convolution $p_{1\times 1}$, consequently resulting in a reduction in computational costs.

\begin{equation}
\label{eq:bcf}
    \hat{x}^{(f)} = \sum\nolimits_{j=1}^{T}(x_{ij}^{(f)}+v_{j}^{(f)})
\end{equation}

\begin{equation}
\label{eq:bct}
    \hat{x}^{(t)} = \sum\nolimits_{i=1}^{F}(x_{ij}^{(t)}+v_{i}^{(t)})
\end{equation}

Finally, the feature maps coming from the time and frequency separate paths, $\hat{x}^{(f)}$ and $\hat{x}^{(t)}$, are concatenated together by channels to get the output feature $y$ in $\mathbb{R}^{C '\times F \times T}$.

\begin{table}
\centering
\begin{tabular}{c|c|c|c|c}
\hline
Output Shape& Architecture& $k$& $s$& $p$\\
\hline\hline
$1,F,T$& Input& -& -& -\\
$C/2,F/2,T/2$& ConvBnRelu& 3& 2& 1\\
$2C,F/4,T/4$& ConvBnRelu, $g$=$C/2$& 3& 2& 1\\
$C,F/4,T/4$& TF-SepCovs $\times2$& -& -& -\\
$C,F/8,T/8$& MaxPool& 2& 2& 0\\
$1.5C,F/8,T/8$& TF-SepCovs $\times2$& -& -& -\\
$1.5C,F/16,T/16$& MaxPool& 2& 2& 0\\
$2C,F/16,T/16$& TF-SepCovs $\times2$& -& -& -\\
$2.5C,F/16,T/16$& TF-SepCovs $\times3$& -& -& -\\
$10,F/16,T/16$& Conv& 1& 1& 0\\
$10,1,1$& Avgpool& -& -& -\\
\hline
\end{tabular}
\caption{\label{tab:architecture}\textbf{Architecture of TF-SepNet}. $C$, $F$, and $T$ respectively represent channels, frequency bins, and time clips of feature maps. $k$, $s$, $p$ and $g$ separately denote kernel size, stride, padding and group.}
\end{table}

\subsection{Network Architecture}
\label{ssec:network}
Time-Frequency Separate Network (TF-SepNet) is a deep CNN architecture tailored specifically for ASC tasks, which aims at a balance between model complexity and classification accuracy. The model architecture is depicted in Table \ref{tab:architecture}, and a detailed explanation is provided below.

The input spectrogram is in $\mathbb{R}^{1 \times F \times T}$. The TF-SepNet starts from two $3 \times 3$ convolution kernels with 2 strides for initial downsampling. After that, a total of 9 TF-SepConvs described in Section \ref{ssec:tf-sep} are followed. In addition, two $2 \times 2$ maxpooling layers with 2 strides are inserted between the TF-SepConvs for intermediate downsampling, enabling the network to capture more high-level representations. The last phase involves a $1 \times 1$ convolutional layer followed by a global average pooling, allowing the model to obtain multi-class probabilities as the output. Moreover, adaptive residual normalization \cite{Cai2023a} is also plugged in after the initial downsampling block and after every block of TF-SepConvs.

The channel width of TF-SepNet, denoted as $\tau$, serves as a hyperparameter to adjust the complexity of the model \cite{kim21l_interspeech}. By tuning $\tau$, TF-SepNet-$\tau$ can be customized to meet specific needs, ranging from resource-constrained environments to high-performance computing systems.

\begin{table}[t]
    \centering
    \begin{tabular}{l|c c c}
    \hline
    Model& Acc/$\%$& MACs/M& Param/K\\
    \hline \hline
    DCASE Baseline \cite{MartinMorato2022}& 42.9& 29.2& 46.5\\
    \hline
    BC-ResNet-40 \cite{kim21l_interspeech}& 57.1& 17.2& 88.1\\
    BC-Res2Net-40 \cite{Lee2022}& 59.1& 17.2& 85.8\\
    TF-SepNet-40 (ours)& \textbf{60.0}& \textbf{7.0}& \textbf{53.4}\\
    \hline
    BC-ResNet-80 \cite{kim21l_interspeech}& 58.4& 45.8& 315.0\\
    BC-Res2Net-80 \cite{Lee2022}& 59.6& 42.7& 307.0\\
    TF-SepNet-80 (ours)& \textbf{61.6}& \textbf{24.2}& \textbf{196.7}\\
    \hline
    \end{tabular}
    \caption{\textbf{Evaluation results} on the test set of TAU Urban Acoustic Scene 2022 Mobile development dataset \cite{Mesaros2018}. \textbf{Acc} denotes the top-1 accuracy on test set. \textbf{MACs} (Multiply-Accumulate Operations) indicates the computational costs per inference. \textbf{Param} represents the number of parameters.}
    \label{tab:performance}
\end{table}

\section{Experiments}
\label{sec:exp}

\subsection{Dataset and Preprocessing} 
The experiments are conducted with the TAU Urban Acoustic Scene 2022 Mobile development dataset \cite{Mesaros2018}, which is a widely recognized benchmark for the task of low-complexity ASC. We follow the official training/test split of 7:3. The recordings in this dataset were captured by various mobile devices across multiple cities worldwide, introducing challenges to the generalization ability of ASC model.

All audio segments are down-sampled to 32kHz \cite{Schmid2022}. Short-Time Fourier Transform (STFT) is employed to extract time-frequency representations, with a window size of 3072 and a hop size of 500. Following the STFT, a Mel-scaled filter bank with 256 frequency bins and 4096 FFT is applied to transform the spectrograms into a Log-Mel spectrograms.

\subsection{Training Setup}
The proposed model is trained for 100 epochs using the Adam optimizer with default settings in the Pytorch environment. The batch size is set to 32. A warmup \cite{goyal2017accurate} strategy is introduced for fast convergence and stable training. The learning rate is linearly increased from 0 to 0.01 over the first five epochs, subsequently decayed to 0 for the remaining epochs using cosine annealing \cite{loshchilov2017sgdr}. In addition, Mixup \cite{zhang2018mixup} and Freq-MixStyle \cite{Schmid2022} are introduced to improve the overfitting problem. The $\alpha$ of Mixup is set to 0.3 and the $\alpha$ and $p$ of Freq-MixStyle are respectively set to 0.3 and 0.7.

\begin{figure}
    \centering
    \includegraphics[width=\linewidth]{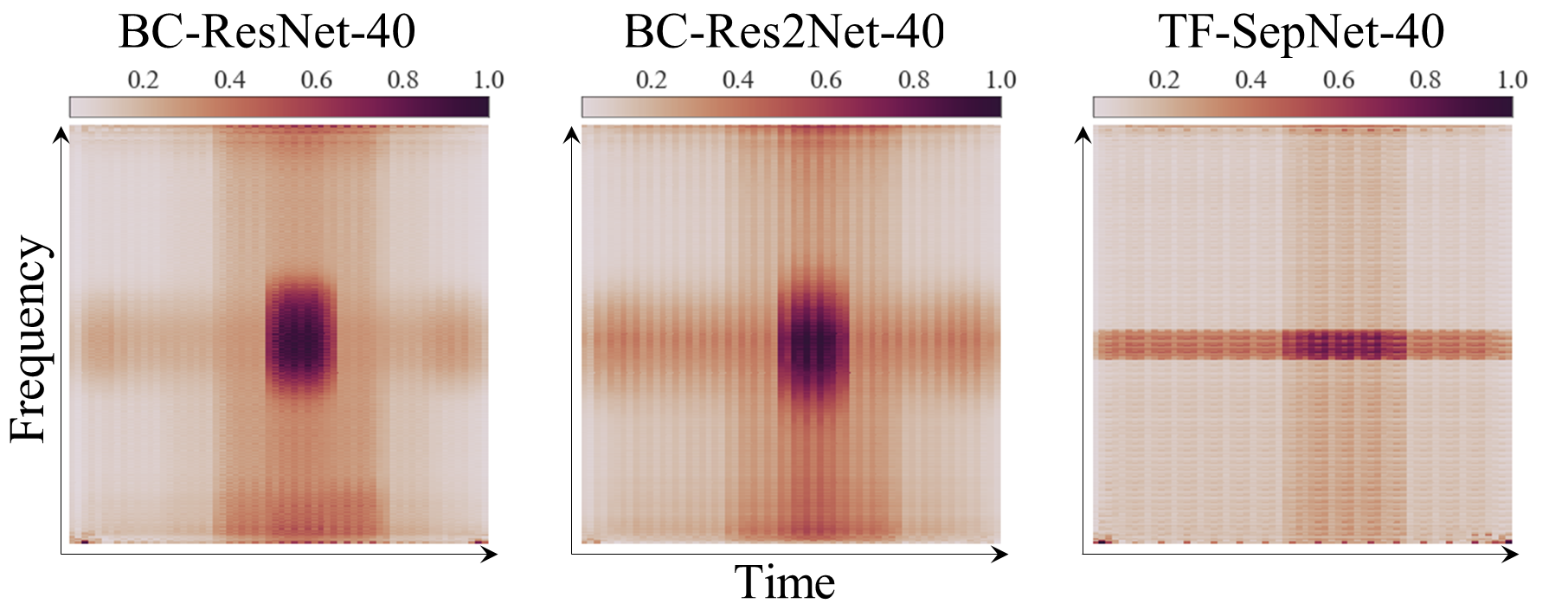}
    \caption{\textbf{Visualization of Effective Receptive Fields (ERF)}. The color intensity at each point signifies the contribution score of the corresponding pixel in the input image to the central point of the feature map generated by the final layer. A broader and darker region indicates a larger ERF.}
    \label{fig:erf}
\end{figure}

\subsection{Results}
Two consecutive-kernels-based CNNs, BC-ResNet \cite{kim21l_interspeech} and BC-Res2Net \cite{Lee2022}, are chosen for comparison with TF-SepNet due to the same scaling mechanism and similar architectures. Notably, BC-ResNet ranked the first in the DCASE2021 Challenge and BC-Res2Net won the second place in 2022. As depicted in Table \ref{tab:performance}, with the channel width $\tau$ being set to 40, TF-SepNet-40 achieves the highest accuracy with 59$\%$ lower computational costs than BC-ResNet-40 and BC-Res2Net-40. Moreover, the parameter number of TF-SepNet-40 is 39$\%$ smaller than BC-ResNet-40 and 38$\%$ less than BC-Res2Net-40. Even with $\tau$ set to 80, TF-SepNet still shows superiority on the performance and efficiency.

The underlying reason for such improvements observed in TF-SepNet is further investigated with the Effective Receptive Field (ERF) \cite{luo2016understanding}. Following \cite{ding2022scaling}, we conducted visualizations and statistical analyses of ERF for BC-ResNet-40, BC-Res2Net-40 and TF-SepNet-40. As illustrated in Fig. \ref{fig:erf}, BC-ResNet and BC-Res2Net show concentrated high-intensity regions around the central point, indicating limited ERF coverage for outer points. In contrast, TF-SepNet exhibited more uniformly distributed high-intensity regions along both the time and frequency dimensions. Table \ref{tab:erf} further corroborates these findings, revealing that TF-SepNet consistently achieved larger ERFs compared to BC-ResNet and BC-Res2Net, irrespective of whether $t$ was set to 20$\%$, 30$\%$, or 50$\%$. With a larger ERF, TF-SepNet is able to capture more time-frequency features from the input audio spectrograms, leading to an improved performance.

\subsection{Ablation Study}
Table \ref{tab:ablation} shows the impacts of key components within the TF-SepNet architecture. The shuffle unit slightly enhances the accuracy without introducing additional parameters or computation by establishing information flow between channel groups. `w/o freq/temp path' means the channels are not split into two halves in the TF-SepConvs module and the entire frequential or temporal path described in Equation (\ref{eq:avgf}) or (\ref{eq:avgt}) is removed. Specially, due to the asymmetrical shape of input spectrogram, with $F$=256 and $T$=64 in our experiments, the frequency and time path inherently incur different computational costs. The accuracy witnesses drastic drops and the system complexity increases regardless of whether the frequential or temporal path is eliminated. The findings underscore the vital contribution of combining information flows in both the time and frequency domains to enhance model performance for ASC. Lastly, the adaptive residual normalization (AdaResNorm) also plays an crucial role, giving rise to 1.5$\%$ accuracy gain by introducing only 2$\%$ parameters.

\begin{table}
    \centering
    \begin{tabular}{l|c c c}
    \hline
    \multirow{2}{*}{Model}& $t=20\%$& $t=30\%$& $t=50\%$\\
    & $r$& $r$& $r$ \\
    \hline \hline
    BC-ResNet-40 \cite{kim21l_interspeech}& 9.6$\%$& 17.3$\%$& 39.3$\%$\\
    BC-Res2Net-40 \cite{Lee2022}& 9.9$\%$& 18.9$\%$& 39.8$\%$\\
    TF-SepNet-40 (ours)& \textbf{13.9$\%$}& \textbf{22.5$\%$}& \textbf{43.8$\%$}\\
    \hline
    \end{tabular}
    \caption{\textbf{Statistical Analysis of ERF}. The threshold ($t$) denotes the selected proportion of all pixel contributions. The high-contribution area ratio ($r$) is the percentage of area around the central point contributing to corresponding $t$ of all pixel contributions. A larger $r$ indicates a more uniform distribution of pixel contributions, in other words, a larger ERF.}
    \label{tab:erf}
\end{table}

\begin{table}
    \centering
    \begin{tabular}{l|c c c}
    \hline
    Model& Acc/$\%$& MACs/M& Param/K\\
    \hline \hline
    TF-SepNet-40& \textbf{60.0}& 7.03& 53.4\\
    \hspace{8pt}w/o shuffle& 59.5& 7.03& 53.4\\
    \hspace{8pt}w/o freq path& 56.7& 7.80& 80.0\\
    \hspace{8pt}w/o temp path& 57.5& 6.96& 80.0\\
    \hspace{8pt}w/o AdaResNorm& 58.5& 7.03& 52.3\\
    \hline
    \end{tabular}
    \caption{\textbf{Ablation Study.} `w/o' means without corresponding component from TF-SepNet-40.}
    \label{tab:ablation}
\end{table}

\section{Conclusion}
\label{sec:conclusion}
This paper addresses the crucial challenge of achieving low system complexity in ASC tasks by introducing TF-SepNet, a novel CNN architecture that leverages separate 1D kernels to independently capture features of audio spectrograms in the time and frequency dimensions. The experimental results show that TF-SepNet achieves better accuracy with lower complexity than the state-of-the-arts that employ consecutive kernels. A study of effective receptive field (ERF) further demonstrates the benefits of separate kernels in the ASC task. For future works, we believe the separate kernels present a huge potential in other audio pattern recognition tasks.

\section{Acknowledgement}
This project is supported partly by the National Natural Science Foundation of China (No: 62001038) and Gusu Innovation and Entrepreneurship Leading Talents Programme (No: ZXL2022472). We also thank Dr. Yin Cao and Dr. Jimin Xiao for their valuable advice on this paper.

%\vfill
%\pagebreak

\bibliographystyle{IEEEbib}
\bibliography{reference}

\begin{thebibliography}{10}

\bibitem{barchiesi2015acoustic}
Daniele Barchiesi, Dimitrios Giannoulis, Dan Stowell, and Mark~D Plumbley,
\newblock ``Acoustic scene classification: Classifying environments from the
  sounds they produce,''
\newblock {\em IEEE Signal Processing Magazine}, vol. 32, no. 3, 2015.

\bibitem{MartinMorato2022}
Irene Mart\'{i}n-Morat\'{o}, Francesco Paissan, Alberto Ancilotto, Toni
  Heittola, Annamaria Mesaros, Elisabetta Farella, Alessio Brutti, and Tuomas
  Virtanen,
\newblock ``Low-complexity acoustic scene classification in {DCASE} 2022
  {C}hallenge,''
\newblock in {\em Proceedings of the Detection and Classification of Acoustic
  Scenes and Events (DCASE) Workshop}, 2022.

\bibitem{abesser2020review}
Jakob Abe{\ss}er,
\newblock ``A review of deep learning based methods for acoustic scene
  classification,''
\newblock {\em Applied Sciences}, vol. 10, no. 6, 2020.

\bibitem{Koutini2019}
Khaled Koutini, Hamid Eghbal-zadeh, and Gerhard Widmer,
\newblock ``Receptive-field-regularized {CNN} variants for acoustic scene
  classification,''
\newblock in {\em Proceedings of the Detection and Classification of Acoustic
  Scenes and Events (DCASE) Workshop}, 2019, pp. 124--128.

\bibitem{Hu2021}
Hu~Hu, Chao-Han~Huck Yang, Xianjun Xia, Xue Bai, Xin Tang, Yajian Wang, Shutong
  Niu, Li~Chai, Juanjuan Li, et~al.,
\newblock ``A two-stage approach to device-robust acoustic scene
  classification,''
\newblock in {\em Proceedings of the IEEE International Conference on
  Acoustics, Speech and Signal Processing (ICASSP)}, 2021, pp. 845--849.

\bibitem{Cho2019}
Janghoon Cho, Sungrack Yun, Hyoungwoo Park, Jungyun Eum, and Kyuwoong Hwang,
\newblock ``Acoustic scene classification based on a large-margin factorized
  {CNN},''
\newblock in {\em Proceedings of the Detection and Classification of Acoustic
  Scenes and Events (DCASE) Workshop}, 2019, pp. 45--49.

\bibitem{kim21l_interspeech}
Byeonggeun Kim, Simyung Chang, Jinkyu Lee, and Dooyong Sung,
\newblock ``{Broadcasted Residual Learning for Efficient Keyword Spotting},''
\newblock in {\em Proceedings of the Conference of the International Speech
  Communication Association (INTERSPEECH)}. ISCA, 2021, pp. 4538--4542.

\bibitem{Lee2022}
Joo-Hyun Lee, Jeong-Hwan Choi, Pil~Moo Byun, and Joon-Hyuk Chang,
\newblock ``Multi-scale architecture and device-aware data-random-drop based
  fine-tuning method for acoustic scene classification,''
\newblock in {\em Proceedings of the Detection and Classification of Acoustic
  Scenes and Events (DCASE) Workshop}, 2022.

\bibitem{Phan2022}
Duc~H Phan and Douglas~L Jones,
\newblock ``Low-complexity acoustic scene classification using time frequency
  separable convolution,''
\newblock {\em Electronics}, vol. 11, no. 17, 2022.

\bibitem{Zhang2018}
Xiangyu Zhang, Xinyu Zhou, Mengxiao Lin, and Jian Sun,
\newblock ``{ShuffleNet}: An extremely efficient convolutional neural network
  for mobile devices,''
\newblock in {\em Proceedings of the IEEE Conference on Computer Vision and
  Pattern Recognition (CVPR)}, 2018, pp. 6848--6856.

\bibitem{Mu2021}
Wenjie Mu, Bo~Yin, Xianqing Huang, Jiali Xu, and Zehua Du,
\newblock ``Environmental sound classification using temporal-frequency
  attention based convolutional neural network,''
\newblock {\em Scientific Reports}, vol. 11, no. 1, 2021.

\bibitem{Mesaros2018}
Annamaria Mesaros, Toni Heittola, and Tuomas Virtanen,
\newblock ``A multi-device dataset for urban acoustic scene classification,''
\newblock in {\em Proceedings of the Detection and Classification of Acoustic
  Scenes and Events (DCASE) Workshop}, 2018, pp. 9--13.

\bibitem{luo2016understanding}
Wenjie Luo, Yujia Li, Raquel Urtasun, and Richard Zemel,
\newblock ``Understanding the effective receptive field in deep convolutional
  neural networks,''
\newblock {\em Advances in Neural Information Processing Systems}, vol. 29,
  2016.

\bibitem{Cai2023a}
Yiqiang Cai, Minyu Lin, Chenyang Zhu, Shengchen Li, and Xi~Shao,
\newblock ``{DCASE}2023 task1 submission: Device simulation and time-frequency
  separable convolution for acoustic scene classification,''
\newblock Tech. {R}ep., Detection and Classification of Acoustic Scenes and
  Events (DCASE) Challenge, 2023.

\bibitem{Schmid2022}
Florian Schmid, Shahed Masoudian, Khaled Koutini, and Gerhard Widmer,
\newblock ``{CP-JKU} submission to {DCASE22}: Distilling knowledge for
  low-complexity convolutional neural networks from a patchout audio
  transformer,''
\newblock Tech. {R}ep., Detection and Classification of Acoustic Scenes and
  Events (DCASE) Challenge, 2022.

\bibitem{goyal2017accurate}
Priya Goyal, Piotr Doll{\'a}r, Ross Girshick, Pieter Noordhuis, Lukasz
  Wesolowski, Aapo Kyrola, Andrew Tulloch, Yangqing Jia, and Kaiming He,
\newblock ``Accurate, large minibatch {SGD}: Training imagenet in 1 hour,''
\newblock {\em arXiv preprint arXiv:1706.02677}, 2017.

\bibitem{loshchilov2017sgdr}
Ilya Loshchilov and Frank Hutter,
\newblock ``{SGDR}: Stochastic gradient descent with warm restarts,''
\newblock in {\em Proceedings of the International Conference on Learning
  Representations (ICLR)}, 2017.

\bibitem{zhang2018mixup}
Hongyi Zhang, Moustapha Cisse, Yann~N. Dauphin, and David Lopez-Paz,
\newblock ``mixup: Beyond empirical risk minimization,''
\newblock in {\em Proceedings of the International Conference on Learning
  Representations (ICLR)}, 2018.

\bibitem{ding2022scaling}
Xiaohan Ding, Xiangyu Zhang, Jungong Han, and Guiguang Ding,
\newblock ``Scaling up your kernels to 31x31: Revisiting large kernel design in
  {CNN}s,''
\newblock in {\em Proceedings of the IEEE/CVF Conference on Computer Vision and
  Pattern Recognition (CVPR)}, 2022, pp. 11963--11975.

\end{thebibliography}

\end{document}